\documentclass[lettersize,journal]{IEEEtran}
\usepackage{amsmath,amsfonts}
\usepackage{algorithmic}
\usepackage{array}
\usepackage[caption=false,font=normalsize,labelfont=sf,textfont=sf]{subfig}
\usepackage{textcomp}
\usepackage{stfloats}
\usepackage{url}
\usepackage{verbatim}
\usepackage{graphicx}
\usepackage{color}
\usepackage[dvipsnames]{xcolor}
\usepackage{booktabs}
\usepackage[para]{threeparttable}
\usepackage{soul}

\def\BibTeX{{\rm B\kern-.05em{\sc i\kern-.025em b}\kern-.08em
    T\kern-.1667em\lower.7ex\hbox{E}\kern-.125emX}}
\usepackage{balance}

\begin{document}

\title{Scalable Multiple Electron Transport Architectures for Feedback Traceable Current Sources}

\author{
Agustin Javier Lapi\IEEEauthorrefmark{1}\IEEEauthorrefmark{2}\IEEEauthorrefmark{3},
Guillermo Fernandez Moroni\IEEEauthorrefmark{3},
Fernando Chierchie\IEEEauthorrefmark{4}\IEEEauthorrefmark{2},
Fabricio Alcalde Bessia\IEEEauthorrefmark{5}\IEEEauthorrefmark{6},
Miqueas Ezequiel Gamero\IEEEauthorrefmark{4}\IEEEauthorrefmark{2}\IEEEauthorrefmark{3},
Brenda Aurea Cervantes Vergara\IEEEauthorrefmark{3},
Blas Junior Irigoyen Gimenez\IEEEauthorrefmark{4}\IEEEauthorrefmark{2},
Eduardo Paolini\IEEEauthorrefmark{4}\IEEEauthorrefmark{2},
Claudio Rodrigo Chavez Blanco\IEEEauthorrefmark{4}\IEEEauthorrefmark{3},
Juan Estrada\IEEEauthorrefmark{3},
Javier Tiffenberg\IEEEauthorrefmark{3}
\\[1ex]

\IEEEauthorrefmark{1}Department of Astronomy and Astrophysics, University of Chicago, Chicago, IL, USA\\
\IEEEauthorrefmark{2}Instituto de Inv. en Ing. Eléctrica ``Alfredo Desages'' (IIIE), CONICET and Universidad Nacional del Sur (UNS), Bahía Blanca, Argentina\\
\IEEEauthorrefmark{3}Fermi National Accelerator Laboratory, Batavia IL, United States\\
\IEEEauthorrefmark{4}DIEC - Universidad Nacional del Sur, Bahia Blanca, Argentina\\
\IEEEauthorrefmark{5}Instituto Balseiro, San Carlos de Bariloche, Argentina\\
\IEEEauthorrefmark{6}Instituto de Nanociencia y Nanotecnología INN (CNEA-CONICET), San Carlos de Bariloche, Argentina\\
}

\maketitle

\begin{abstract}
We present a scalable electron-counting current cell based on a floating gate architecture operated as a Multiple Electron Transport device (MET). The system enables controlled injection, transport, and precise quantification of discrete charge packets using the sensor non-destructive multiple readout capability. We repurposed the charge-injection technique, jointly with the electron-resolution capability, for a controlled generation of quantized charge packets for an electron-traceable current source. 

Scalable architectures based on parallel and series multi-amplifier configurations are explored. Experimental results confirm noise reduction following the square root of the number of independent measurements and demonstrate stable, programmable output current. This approach provides a compact and scalable platform for electron-counting current sources and precision quantum metrology applications.

\end{abstract}

\begin{IEEEkeywords}
single-electron, scalable architecture, compact current cell, multiple electron transport device
\end{IEEEkeywords}

\section{Introduction}
\label{sec:intro}  
The generation of ultra-low currents with metrological accuracy has become relevant following the redefinition of the ampere in terms of the elementary charge \cite{BIMP2019}. A direct realization of the ampere requires a known number of electrons flowing per unit of time, establishing traceability to the elementary particle. Over the past decades, quantum-based current sources have been studied, most notably single-electron transistors (SETs), single-electron pumps, and turnstile devices, which exploit tunneling through nanostructures to generate currents in the form of $I=fe$, where $f$ is the single-electron pumping rate \cite{yamahata2014gigahertz, Bae_2020, single_level_turnstile}. These methods control the transfer of individual electrons at high repetition rates, producing well-defined current levels in the femtoampere to picoampere range. However, as reviewed in \cite{Kaneko2024}, single-electron and other quantum-based current standards require complex nanofabrication, operation at cryogenic temperature, and high-frequency control (in the GHz order), while their scalability and achievable range remain limited. Alternative room-temperature approaches offer operational simplicity but lack direct electron-traceability \cite{okazaki_2022}. More recently, CCD quantum image sensors, which achieve single-electron resolution and can be directly traced to the elementary charge, have been explored. In these devices, charge generated by optical illumination and collected in the sensor is read out at a known rate to define an equivalent current \cite{Gamero_2025}.

Sensors with single electron resolution are emerging in the field as a conceptual alternative in the current generation, enabling the direct measurement of larger charge packets rather than relying on controlled transport of individual electrons \cite{Tiffenberg:2017aac}. In this framework, a current is defined by the draining of multiple charge packets, measured with electron-level precision, at a given frequency. This gives place to the Multiple Electron Transport devices (METs), which can trace to the electron charge even when packets contain many electrons.
In this direction, the Skipper-CCDs have a floating gate output stage enabling multiple non-destructive measurements of the same charge packet, achieving single-electron resolution. A proof of concept of this idea was recently published by \cite{Gamero_2025}, where a Skipper-CCD was used to collect light from an optical source, measure the resulting charge packets in the tens to thousand range, and drain them, thus implementing a self-referenced electron-based current source. This approach reduces the need to pump single electrons at GHz rates at very low cryogenic temperatures.

However, current demonstrations relied on optical illumination to generate charge packets. Poisson statistics, sensitivity to illumination stability, and the relation between exposure time and charge magnitude intrinsically limit this method. In this work, we use a controlled charge-injection technique to generate charge packets directly within the CCD. Electronic charge injection, has been historically used in CCDs as a memory-write mechanism, in signal processing and more recently for radiation-damage mitigation and trap characterization in space applications \cite{nobel_invention_ccd, linearity_sequin, ctiCorrectionWithChargeInjection}. This technique provides more control, reproducibility, and independence from optical sources. As shown in earlier studies \cite{ChargeInjectionNoise,Koch_ChargeInjectionDevice}, the standard deviation associated with charge injection can be significantly lower than photon-generated charge for large packets. In this article, we repurpose this technique for a new application: the controlled generation of quantized charge packets for electron-traceable current sources.

We present experimental results obtained using a Skipper-CCD operated with the reuse of the charge-injection technique. The injected charge packets, measured with single-electron resolution, are drained to produce an output current. This approach enables a novel method for absolute charge and current calibration. By directly measuring injected charge packets from the single-electron level up to 2500 electrons, we establish a calibration scheme that is intrinsically traceable to the elementary charge and spans an unprecedented dynamic range within a single device. These results highlight the potential of floating gate based architectures as scalable, self-calibrated METs.
 
Motivated by these experimental results, we propose extending the Skipper-CCD concept, as a MET device, toward a compact basic current cell capable of generating a quantized current unit without optical illumination. The compact design facilitates scalability, enabling the generation of larger current signals through arrayed structures. We explore three architectures based on floating amplifiers, single-amplifier, series and parallel, along with their expected performance, scalability, output current, and achievable current range.

The paper is organized as follows: next section explores the concept of a current cell in open-loop and closed-loop configurations. The open-loop diagram is extended to three cell architectures, where a charge quantizer block is incorporated, with a noise and noise-to-signal description. Section \ref{sec:simulation_results} presents simulation results for the three architectures explored. Section \ref{sec:charge_transfer_mechanisms} presents the charge transfer mechanisms of a skipper-CCD, drain in and drain out of the sequential memory for both, for charge injection and its standard operation, respectively. In Section \ref{sec:experimental_results} the configuration used with a CCD to perform experiments, along with the setup and experimental results are presented. Extensive measurements include a charge-injection curve, absolute calibration of charge packets up to $2500$e$^-$, and temporal current stability. In the section \ref{sec:layout_basic_cell}, a layout for the proposed current cell is presented. Finally, in Section \ref{sec:conclusions}, the conclusions and future work are addressed.

\section{Multiple-cell architectures}
\label{sec:multiple-cell_architecture_and_performance}
\begin{figure}[htbp]
    \centering

    \subfloat[]{
        \includegraphics[width=\linewidth]
            {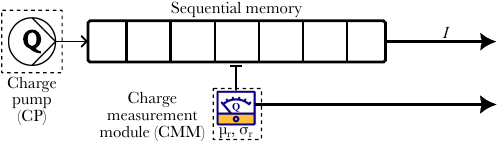}
        \label{fig:objective_scheme_a}
    }

    \subfloat[]{
        \includegraphics[width=\linewidth]
            {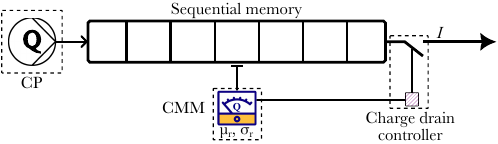}
        \label{fig:objective_scheme_b}
    }

    \caption{(a) is an open-loop Basic current cell conceptual sketch. The open-loop is extended to a closed-loop by adding a charge drain controller block in (b).}
    \label{fig:objective_scheme}
\end{figure}

In this section, two simplified block diagrams defining current cell architectures are presented in Figure \ref{fig:objective_scheme}. The Figure \ref{fig:objective_scheme_a} diagram illustrates an open-loop configuration, while the Figure \ref{fig:objective_scheme_b} diagram shows an extension of the same concept in a closed-loop topology.
In the open-loop scheme, a charge pump (CP) delivers discrete charge packets into a sequential charge analog memory, analogous to the serial register of a CCD. The sequential memory stores the multi-electron charge packets generated by the CP. The CP charge injection mechanism is a naturally random process characterized by a mean value $\mu_r$ and a statistical dispersion $\sigma_r$. 
After passing through the memory, the charge is read out by a charge-measurement module (CMM), known as the output stage. The CMM block uses a floating gate coupled to a MOS amplifier output stage, allowing for multiple non-destructive measurements of the same charge packet, reaching the sub-electron noise regime at a time penalty cost. 
In this configuration, the output current ($I$) is defined by the rate at which the charge packets are drained from the system. The current source delivers a current together with its precise charge measurement using the CMM. In this case, external equipment can use the current supplied and compare its value with the measurements provided by the current source itself.
The closed-loop architecture, depicted in the Figure \ref{fig:objective_scheme_b}, extends this concept by incorporating a charge drain controller block. This block serves as a feedback mechanism to hold a desired current flowing from the current source. This controller uses information from the CMM to dynamically adjust the drain rate, thus regulating the output current in real time. The drain controller block falls outside the scope of this work.

\begin{figure}[htbp]
    \centering

    \subfloat[]{
        \includegraphics[width=0.9\linewidth]
            {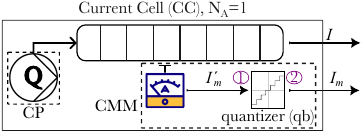}
        \label{fig:generic_architectures_a}
    }

    \subfloat[]{
        \includegraphics[width=\linewidth]
            {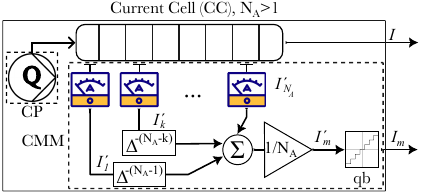}
        \label{fig:generic_architectures_b}
    }

    \subfloat[]{
        \includegraphics[width=0.9\linewidth]
            {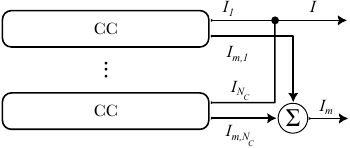}
        \label{fig:generic_architectures_c}
    }

    \caption{Schematic representation of the proposed architecture.
    In (a), the current cell is formed by a charge measurement block
    (CMM), a charge-pump (CP) block, and a memory.
    The cell is extended in (b), where the CMM can measure the current
    with different amplifiers, align and average these measurements to
    obtain a more precise mean current.
    Many of these CCs can be configured in parallel to achieve an
    increased current value, as presented in (c).}
    \label{fig:generic_architectures}
\end{figure}

This sequential memory structure can be used to develop a \'basic cell\' of which three architectures are presented, aiming to increase the current magnitude and enhance accuracy. These strategies could be implemented using either of the previous topologies. For simplicity, only the open-loop current cell shown in Figure~\ref{fig:objective_scheme_a} is considered and extended to three architectures of increasing complexity, as illustrated in Figure~\ref{fig:generic_architectures}. The first (a) corresponds to the most compact basic cell, followed by a more general architecture (b) that employs multiple CMMs along the serial register to perform independent measurements of the same charge packet, of which (a) is a special case. The third (c) further extends this approach by operating multiple cells in parallel to increase the output current level.

\subsection{Basic cell}
For the basic current cell showed in Figure \ref{fig:generic_architectures_a}, a CP source models the charge injection into the serial register. At the end of the serial register, an output current $I$ is obtained as a result of the charge transfer process. The charge flowing through the register is measured by the CMM and, using the pixel transfer time, is converted into an equivalent measured current $I'_m$.
The current measurement, as we will see later, is implemented with a MOS amplifier which measures the charge of the pixel ($q(n)$) connected to it, where $n$ is the pixel index. The measured current can be defined as $i(n)=c\frac{q(n) + r(n)}{T_{PIX}}$, where $T_{PIX}$ represents the readout period of the pixels and $r(n)$ is the measurement error related to the readout electronic noise of the transistor. Both quantities, $q(n)$ and $r(n)$, are expressed in electron units, and $c$ is the electron elementary charge in coulomb units. 
$r(n)$ follows a Gaussian distribution with zero mean and standard deviation $\sigma$, i.e., $r(n)\sim \mathcal{N}(0,\sigma)$. This uncertainty can be modeled as 
\begin{equation}
\sigma=\frac{\sigma_0}{\sqrt{N_S}},     
\label{eq:sigma_reduction}
\end{equation}
where $\sigma_0$ is the uncertainty of a single measurement of the charge packet and $N_S$ is the number of independent measurements taken from the same charge packet. $\sigma_0$ is usually dominated by the built-in readout amplifier.

A quantizer is included in the extended CMM block to account for the quantized nature of electric charge, converting the measured current $I'_m$ into a discretized measured current $I_m$. The role of the quantizer is to associate the charge accumulated in each pixel with a discrete electron multiplicity.

The impact of the quantizer on the uncertainty of the current measurement depends on the measurement standard deviation, $\sigma$, and can be analyzed as follows. The quantizer is modeled as a function $f$ acting on the charge measurement, $f(q(n)+r(n))$, where $f$ maps the noisy charge value to the nearest integer, assuming no missing codes (electron counts):

\begin{equation}
    f(q(n)+r(n)) = q(n) + f(r(n))
    \label{eq:quantizer}
\end{equation}
$f(r(n))$ represents the integer rounding of the noise component. In the ideal case, if $r(n)=0$, then $f(q(n))=q(n)$, yielding the exact value of the charge packet, with $q(n) \in\mathbf{Z}$.
Therefore, a new random variable $q'(n)$ is defined to represent the quantized version of the noisy charge measurement:
 \begin{equation}
     q'(n) = q(n) + f(r(n)) \text{, with } q(n),f(r(n))\in\mathbf{Z},
 \end{equation}
this holds since $q(n)$ is an integer and the quantizer is invariant under integer translations.

This random variable $q'(n)$ is discrete and has mean value $q(n)$, since $r(n)$ has zero mean. Its possible values are $q(n)+i$, with $i\in\mathbf{Z}$. The probability mass function (PMF) of $q'(n)$ is determined by the probability that the quantized noise term $f(r(n))$ takes a specific integer value, denoted by $z$. As $r(n)$ follows a Gaussian distribution, this probability is obtained by integrating the Gaussian probability density over the interval that maps to the integer $z$ under the rounding operation. That is,
\begin{equation}
\begin{split}    
    P(q'(n)=q(n)+z)&=P(f(r(n))=z)\\&=\int_{z-0.5}^{z+0.5} \frac{\sqrt{N_S}e^\frac{-r^2N_S}{2\sigma_0^2}}{\sqrt{2\pi\sigma_0^2}} \,dr
\end{split}
\end{equation}
This expression yields the full probability distribution of the quantized charge $q'(n)$, from which we can derive its variance and expectation. The variance of $q'(n)$ is entirely due to the quantization of the noise and is given by:
\begin{equation}
\begin{split}
    (\sigma')^2 = \text{Var}(q'(n))&=\text{Var}(f(r(n))) \\&=\sum_{z=-\infty}^{\infty}z^2P(f(r(n))=z)
\end{split}
\end{equation}
A comparison in performance with and without the quantizer block is shown in Table \ref{table:comparison_1serie}. For this example, we assume $\sigma_0=2$e-. The table shows that for initial values of $N_S$ the $\sigma'$ is higher than $\sigma$, as would be expected. However, for higher $N_S$ (low $\sigma$), the quantizer block improves the current measurement precision. 
That would be the case for $\sigma'=0.213<\sigma=0.25$ for $N_S=64$. 
In this regime, the quantizer effectively reduces the uncertainty of the charge measurement for each $T_{PIX}$ period. Although quantization is usually an undesirable effect in signal processing, here it is a natural consequence of the intrinsically discrete nature of electron charge and becomes beneficial by suppressing noise fluctuations below the single-electron level.

\begin{table}[h!]
\begin{center}
    \begin{tabular}{lll}
\hline
$N_S$ & $\sigma [e^-]$ & $\sigma' [e^-]$ \\ \hline
1     & 2                    & 2.02               \\
4     & 1                    & 1.04               \\
16    & 0.5                  & 0.57               \\
64    & 0.25                 & 0.213              \\
256   & 0.125                & 0.00795            \\
400   & 0.1                  & 0.000757           \\ \hline
\end{tabular}
\caption{Comparison of standard deviation of the measured charge packets, expressed in electron units for simplicity, pre and post-quantizer block, at positions 1 and 2 respectively, for different $N_S$ values.}
\label{table:comparison_1serie}
\end{center}
\end{table}

The expressions for both charge and current are equivalent. To convert from charge to current, we must divide by $T_{PIX}$. 
The total output current for this case is 
\begin{equation}
I=\frac{Q\text{e}^-}{T_{PIX}},    
\label{eq:basic_current}
\end{equation}
where $\text{e}^-$ is the elementary particle.
With the previous variance calculation of the charge measurement, we can find the noise-to-signal ratio (NSR) of the output current, which we define as

\begin{equation}
    NSR = \frac{\sqrt{\text{Var}(I_m)}}{I} = \begin{cases}
    \text{if non-quantizer:}\\
    \frac{\sqrt{Var(r(n))/T_{PIX}^2}}{q(n)/T_{PIX}}=\frac{\sigma_0}{q(n)\sqrt{N_S}}  \\
    \text{if quantizer:} \\
    \frac{\sqrt{Var(f(r(n)))/T_{PIX}^2}}{q(n)/T_{PIX}}=\frac{\sqrt{\text{Var}(f(r(n)))}}{q(n)} 
    \end{cases}
    \label{eq:snr_ifquantizer}
\end{equation}

Reducing readout noise through skipper sampling introduces a time penalty, since the total pixel readout time increases with the number of samples, $T_{\mathrm{PIX}} \propto N_S$, thereby reducing the effective output current. This penalty can be mitigated by placing multiple CMM output stages in series along the same sequential memory, allowing the same charge packet to be measured multiple times while moving through the serial register and thus reducing the impact of skipping on both readout time and current level.
\subsection{Basic cell with serial reading}
The second architecture, referred to as series architecture, shown in Figure~\ref{fig:generic_architectures_b}, models this idea, where each amplifier measures the current $I'_1, I'_2, \dots, I'_{N_A}$, where $N_A$ is the number of amplifiers. These current measurements are performed at different time instants, which is modeled by a discrete delay block with coefficient $\Delta^{-(k)}$, with $0\leq k< N_A$. The delayed measurements are added and averaged to obtain the mean current $I'_m$. For this architecture, the total output current is the same as in the previous case, presented in Eq. \ref{eq:basic_current}, and it is repeated for convenience, $I=\frac{Q\text{e}^-}{T_{PIX}}$. However, combining multiple time-separated measurements of the same charge packet improves the precision of the current estimation.  

In this architecture, the CMM is extended to $N_A$ measurement devices (amplifiers), a sum block, and a $1/N_A$ gain block, plus the quantizer of the averaged measurement, and the delay blocks for the branches. The current cell (CC) is composed of a CP, a sequential memory, and a charge measurement module.

If we consider the case without the quantizer block, the independent variances of the same charge packets are averaged; therefore, the standard deviation in this architecture is defined as $\sigma=\frac{\sigma_0}{\sqrt{N_S}\sqrt{N_A}}$, assuming the same $\sigma_0$ for all. We can add the contribution of the multiple amplifiers into the non-quantizer NSR expression from Eq. \ref{eq:snr_ifquantizer}, to obtain 
\begin{equation}
    NSR = \frac{\sigma_0}{I\sqrt{N_S}\sqrt{N_A}}.
    \label{eq:SNR_NS_NA}
\end{equation}
\subsection{Parallel Connection of Basic Cell with serial reading}
Until now, we have decreased the noise-to-signal ratio by adding complexity to the charge-measurement stage of the basic cell. We now introduce, in Fig.~\ref {fig:generic_architectures_c}, a scheme to increase the current level by connecting multiple cells, in which the current outputs of $N_C$ CCs are combined at a single output node, i.e., $I = I_1 + I_2 + \cdots + I_{N_C} = N_C I$.

The individual measured currents are also summed to obtain the total current, 
\begin{equation}
I_m=I_{m,1}+I_{m,2}+\dots+I_{m,N_C}.
\label{eq:paralel_current}
\end{equation}

For $N_C$ independent currents, the variances add, and therefore the standard deviation increases as $\sqrt{N_C}$. 
The resulting uncertainty of the total measured current is $\sigma=\frac{\sigma_0\sqrt{N_C}}{\sqrt{N_S}\sqrt{N_A}}$, and extending the non-quantizer expression of Eq. \ref{eq:snr_ifquantizer}, leads to a more general noise-to-signal ratio 
\begin{equation}
    NSR = \frac{\sigma_0}{I\sqrt{N_S}\sqrt{N_A}\sqrt{N_C}},
    \label{eq:SNR_without_P}
\end{equation}
where $I$ is the current of a single cell.

To conclude the Section, several distinctions are useful for designing a current source based on the proposed architectures. Increasing the number of measurements per amplifier, $N_S$, decreases the noise-to-signal ratio at the expense of reduced output current, since the current flow becomes slower. Increasing the number of amplifiers per basic cell, $N_A$, decreases the NSR without changing the output current level, at the cost of a larger number of instrumented channels in the system. Finally, increasing the number of basic cells, $N_C$, improves both the noise-to-signal ratio and the output current level, at the cost of a larger number of basic cells and instrumented channels.

\section{Architectures performance}
\label{sec:simulation_results}
In the typical case a master device is used to calibrate another device, the calibration procedure would involve injecting the generated current into the device under test, such as an external amplifier. The measurement is naturally associated with a device-dependent integration time $t_{m}$. 
At this stage, the current is generated at the pixel rate $T_{\mathrm{PIX}}$, while the measurement can be performed over the longer integration time $t_m$. During this interval, a total of $P = t_m / T_{\mathrm{PIX}}$ charge packets are drained and measured. Expressed in units of current, the mean signal remains unchanged, whereas the uncertainty associated with the current cell is averaged over $P$ packets, leading to a noise reduction that scales as $1/\sqrt{P}$.
The pixel time is given by $T_{\mathrm{PIX}} = N_S t_o$, where $t_o$ is the single-sample readout time. Increasing $N_S$ therefore increases $T_{\mathrm{PIX}}$, reducing the number of packets $P$ that fit within a fixed measurement interval $t_m$. As a result, although multiple sampling reduces the noise of each individual measurement, it simultaneously decreases the number of collected packets during $t_m$ in the same proportion. For example, doubling $N_S$ doubles the readout time and halves the number of packets collected within $t_m$.
The use of a different measurement time, will modify the general expression of the NSR for the current cell (series and parallel) as follows,
\begin{equation}
    NSR = \frac{I\sqrt{N_S}\sqrt{N_A}\sqrt{N_C}\sqrt{P}}{\sigma_0},
    \label{eq:SNR_with_P}
\end{equation}
where $P$ is the quantity of packets collected during $t_m$, and $I$ is the current of a single cell.

\begin{figure}[t!]
    \centering

    \subfloat[]{
        \includegraphics[
            page=1,
            width=\linewidth,
            trim={0 0 0 0},
            clip
        ]{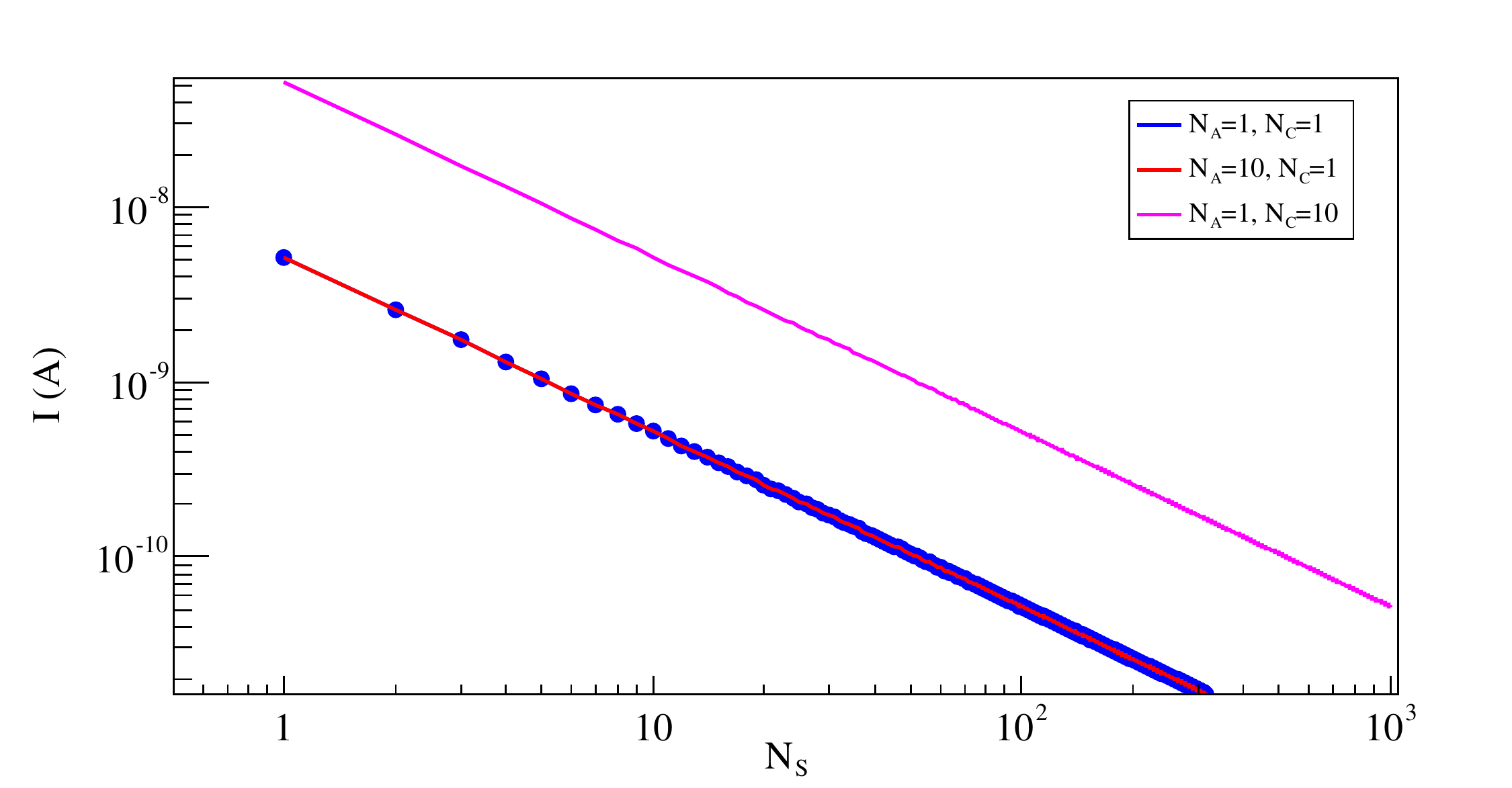}
        \label{fig:signal_optim}
    }

    \subfloat[]{
        \includegraphics[
            page=1,
            width=\linewidth,
            trim={0 0 0 0},
            clip
        ]{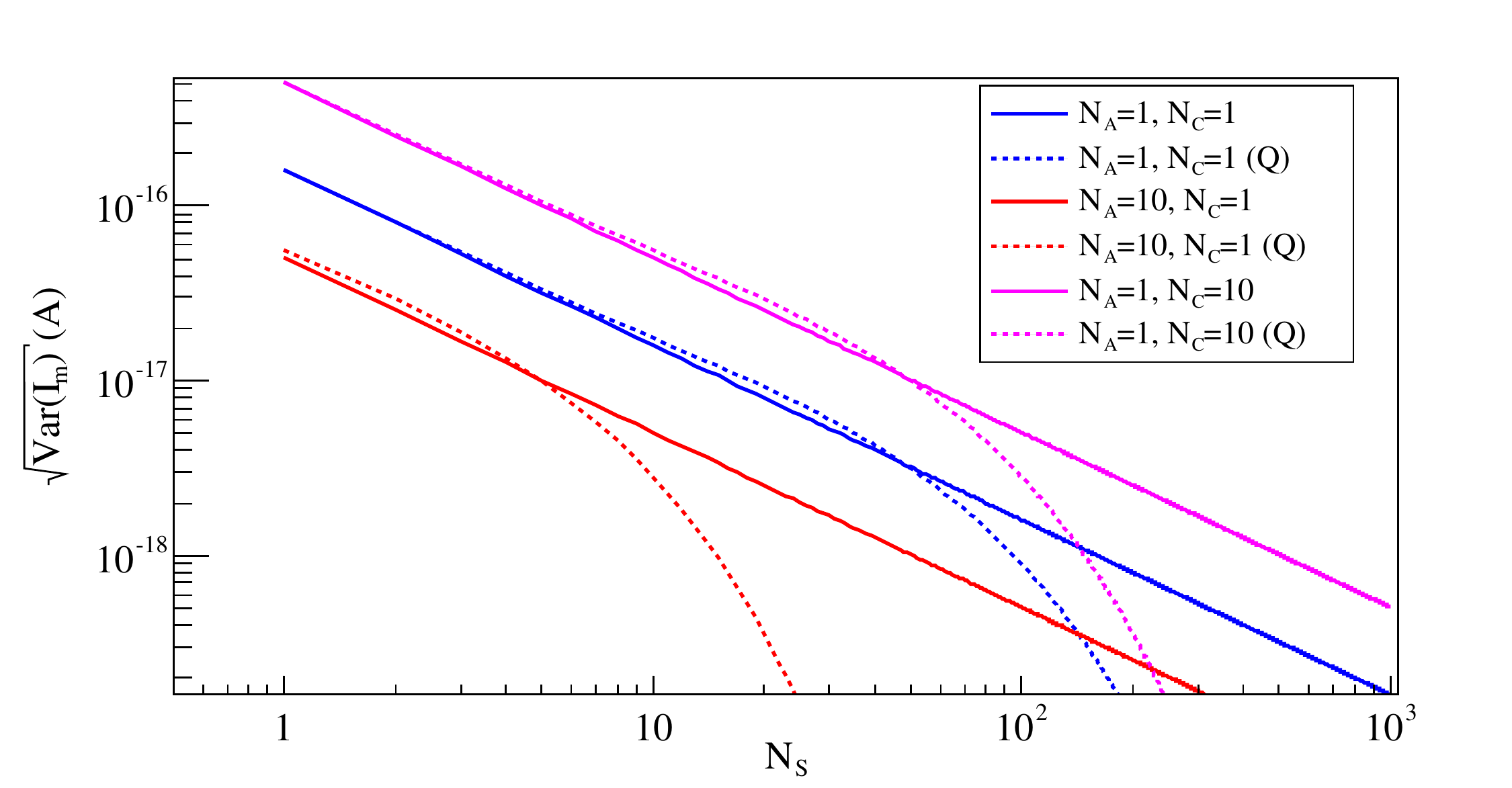}
        \label{fig:noise_optim}
    }

    \subfloat[]{
        \includegraphics[
            page=1,
            width=\linewidth,
            trim={0 0 0 0},
            clip
        ]{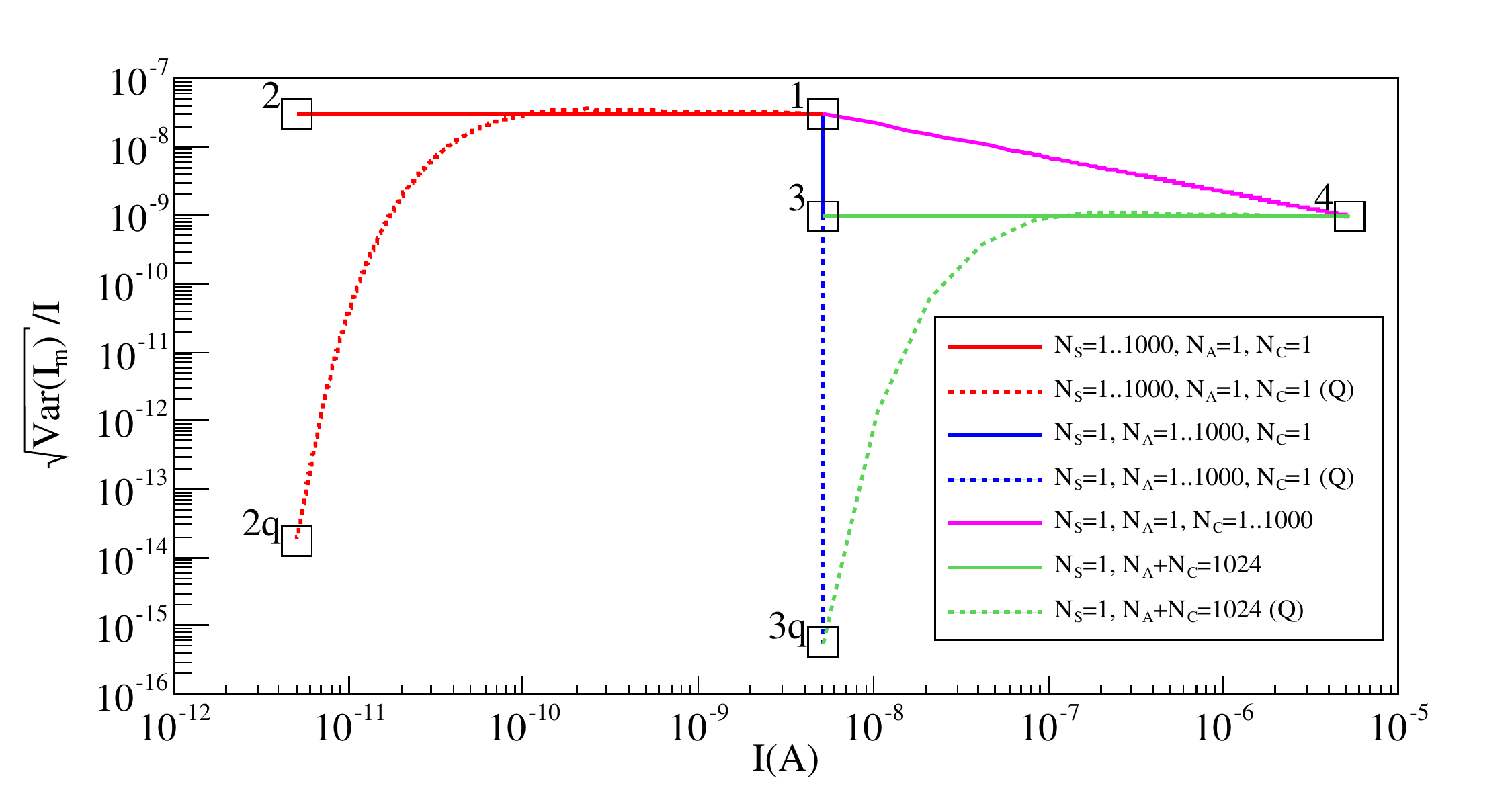}
        \label{fig:relative_error_regions}
    }

    \caption{Comparison of series and parallel architectures. The simulation
    parameters correspond to charge packets of $130\,\mathrm{k}$ electrons,
    a single-sample rate of $250\,\mathrm{kpix/s}$, and a measurement time
    $t_m$ of $1\,\mathrm{s}$.
    (a) Signal as a function of the number of non-destructive measurements
    ($N_S$) for one-amplifier (blue), ten-series (red), and ten-parallel
    (magenta) outputs.
    (b) Noise as a function of $N_S$ for the same configurations. The dashed
    lines indicate the effect of using the quantizer.
    (c) Relative error (error/current) as a function of current, showing the
    behavior for increasing $N_S$, $N_A$, and $N_C$ in red, blue, and
    magenta, respectively. For $N_S$ and $N_A$, the quantizer response is
    also shown by dashed lines.}
    
    \label{fig:error_and_relative_error}
\end{figure}

To evaluate the performance of the different cell architectures, we use as simulation parameters the full-well capacity and readout speed of CCDs of the same technological family as those considered in this work, very well reported. 
These parameters are representative of charge packets of approximately $130$k electrons and a single-sample rate of $250$kpix/s, comparable to those achieved by DECam CCDs \cite{flaugher2015dark, Lapi2022}. This choice defines a realistic and achievable future operating scenario for the proposed current-source architectures. In addition, the measurement time $t_m$ is set to 1 second. 
Following, the cases of a current cell with one amplifier and ten amplifiers in series, as well as the case of ten one-amplifier current cells in parallel, were addressed. The simulation of these architectures was performed with and without the quantizer block.
Figure \ref{fig:signal_optim} shows the current as a function of the $N_S$ value, for the different architectures presented. As the readout time increases, $T_{PIX}$ increases, which is reflected as a lower signal level. The series configuration signal levels are the same, and are represented by blue dots and a red line, respectively, for one or ten amplifiers. Connecting ten cells in parallel increases the signal by a factor of ten, as indicated by the magenta line in the figure. 

For the same configurations, the noise as a function of $N_S$ is also displayed in Figure~\ref{fig:noise_optim}. The difference in noise is well defined, having $N_A$ amplifiers in series diminishes the readout noise by a $N_A$ factor, while adding $N_C$ cells, increases the output noise a $N_C$ factor. The use of the quantizer is indicated in dashed lines. Once the quantized-electron regime is reached, due to its nature, the measured signal becomes discretized, and the noise sharply drops towards zero.

Finally, Figure~\ref{fig:noise_optim} shows the noise-to-signal ratio $\sqrt{\text{Var}(I_m)}/I$ as a function of $I$, for different architecture configurations. Starting from the compact and simplest cell ($N_S=N_A=N_C=1$), and using the same initial conditions for the charge packet and readout rate as the other simulations, the starting point \textbf{1} in the Figure is defined. From here, it is possible to vary the different cell parameters and explore their impact.

An increase in $N_S$ will lower $I$, maintaining $\sqrt{\text{Var}(I_m)}/I$ constant, reflecting the effect of a fixed $t_m$ value, a movement from point \textbf{1} to \textbf{2} is given when $N_S$ varies up to $1000$, this is indicated in red colored line. Adding the quantizer block has the following effect: during the variation of $N_S$ $\sqrt{\text{Var}(I_m)}$ increases slightly for the initial values, however, beyond certain point, this quantity drops fast, resulting in a decreasing curve with an initial bump, indicated by a dashed red line, and arriving to point \textbf{2q}.

On the other hand, according to Eq. \ref{eq:SNR_with_P}, if only $N_A$ is increased, the noise-to-signal ratio becomes smaller. The current in this case remains constant. This is reflected in the figure, where an increase in $N_A$ from 1 to 1000, produces a movement from point \textbf{1} to \textbf{3}, indicated with a blue-colored line. This effect is exacerbated with a quantizer block, indicated with a blue dashed line, final point is \textbf{3q}. 

The third case to analyze, derived from Eq. \ref{eq:SNR_with_P}, is a variation in the number of cells $N_C$. Replicating the cell current increases the total current, and since the noise increases at a lower rate, the noise-to-signal ratio decreases. This is reflected in the figure, when varying $N_C$ from $1$ to $1000$, will increase the signal level and decrease the relation $\sqrt{\text{Var}(I_m)}/I$, going from \textbf{1} to \textbf{4} in the magenta-colored line.

Moreover, we can analyze a combined scenario, where the number of instrumented channels is fixed, resulting in a combination of the number of cells and the number of amplifiers per cell, with all the cells being identical. For a total of 1024 outputs, this combination of $N_A$ and $N_C$ is added in green color. The effect links the final points with a green line (\textbf{3} to \textbf{4}) or a green dashed line (\textbf{3q} to \textbf{4}) for the case with and without the quantizer block, respectively. 

To conclude this section, the simulation results indicates that including the quantizer block has a significant impact when the number of multiple non-destructive readouts is large, i.e. $N_S$ or $N_A$ indiscriminately. For a small number of samples, the quantizer block does not noticeably affect the readout noise. However, as the noise level decreases with increasing samples (and/or in series amplifiers), the system enters a regime in which the benefit of the quantizer becomes substantial. Once the charge has been discretized, further reducing the readout noise is useless. This allows achieving really low current values with high precision.

\section{Single cell operation}
\label{sec:charge_transfer_mechanisms}
In this section, the transfer mechanisms associated with charge injection and normal readout operation of a Skipper-CCD sensor are explained. The sensor is operated in the fully depleted regime. The charge is injected into the sensor through the diode used for charge dumping in normal operation mode.
The Fig. \ref{fig:charge-inj-sequence} is divided into two parts. The left side illustrates the charge injection sequence, while the readout sequence is on the right. The double broken line compress the serial register (array of pixels) between the two stages.

\subsection{Charge injection}
\label{sec:charge-injection-theory}
The method used to inject the charge is known in the literature as the fill and spill technique \cite{Prigozhin_2008, kosonocky1976introducing}. 
In the fill stage, the potential well is filled with charge flowing from the ohmic contact. Then, by changing the gate potential, the excess charge spills back, leaving a defined charge packet whose amount is set by the potential difference between the gates. 

\begin{figure*}
    \centering
    \includegraphics[page=1,width=\textwidth]{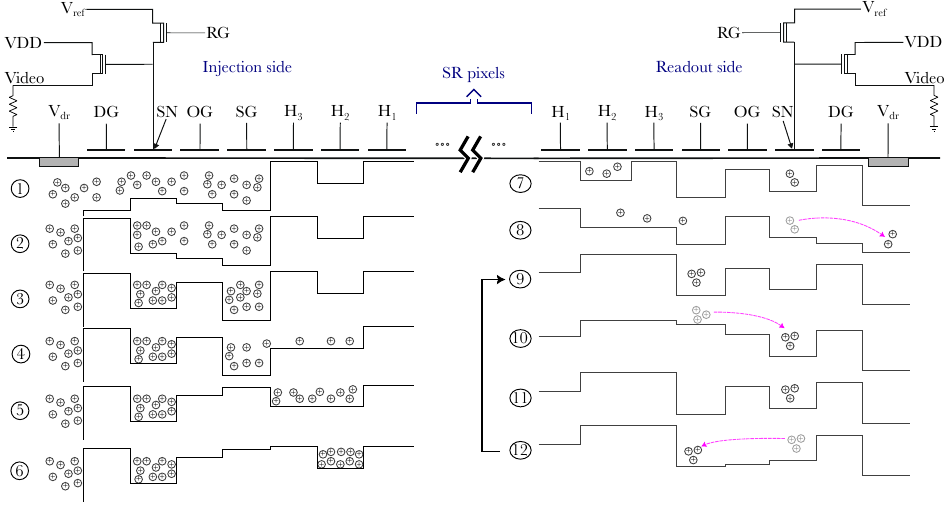}
    \caption{Left side: Charge injection diagram reflecting the fill and spill technique used to move charge from $V_{dr}$ of the injection side to H$_2$ in the readout side. The injection is composed of six steps. The two broken lines compress many 3-phase horizontal phases (pixels) conforming the serial register, which separates the injection from the output side. Right side: horizontal charge transfer and readout operation. In states 7 and 8, the charge is moved from the last pixel position H$_2$ to SG, and the charge beneath the sense node is drained. From states 9 to 12, the readout operation of the charge takes place.
    \label{fig:charge-inj-sequence}}
 \end{figure*}

Figure \ref{fig:charge-inj-sequence} presents a timing diagram illustrating how charge is transferred from the ohmic contact ($V_{dr}$) into the serial register beneath the horizontal gate (H$_2$). The horizontal axis represents the clock states applied to the relevant gates in this process: Dump Gate (DG), Sense Node (SN), Output Gate (OG), Summing Gate (SG), and the three Horizontal phases (H$_3$, H$_2$, H$_1$), over six consecutive states, labeled 1 through 6. 
During the fill stage, DG is clocked low, allowing charge to flow from the ohmic contact to the sense node and accumulate under SG. 
Subsequently, DG is raised, pushing excess charge back and redistributing the collected charge among SN, OG, and SG. In phase 3, the charge packet is transferred beneath SG by clocking OG to a high state.
The following three phases correspond to standard serial register operation, transferring the charge from SG to the first pixel position, under gate H$_2$. By repeating this timing cycle, an amount of charge is injected into the serial register.

 ~%

The amount of charge injected with the fill and spill methodology follows a probability distribution that is bound to the thermal diffusion model, with a more detailed explanation in \cite{ChargeInjectionNoise}. It is interesting to note that the variance in the amount of injected Q in this process is lower than the expected noise in a Poisson process for the same amount of generated charge. 
The fill-and-spill methodology is proposed as an improvement over the proof of concept presented in \cite{Gamero_2025}, where light from an LED was used to generate charge. In contrast, the fill-and-spill approach enables controlled and repeatable charge generation without relying on light injection or the sensor’s active area for charge collection, requiring only the serial register for operation, simplifying the experimental setup. This allows the establishment of a continuous charge flow in which the current source can operate indefinitely under steady-state conditions.

\subsection{Charge readout and drain out}
In the right part of the Figure \ref{fig:charge-inj-sequence}, a time diagram of readout operation of a Skipper-CCDs output stage is shown. 
The control of the charge packet in the output stage is done by the clocks applied to the SG, the OG, the Reset Gate (RG), and the DG. Initially, the injected charge is below H$_2$, from steps 7 to 8, the horizontal clocks are phased to move the charge to SG. At the same time, the last measured pixel charge, which is located in the SN, is removed through $V_{dr}$ by operating the dump gate. Simultaneously, the RG gate activates the Skipper-CCD reset transistor to restore the SN to its reference level. In step 9, H$_3$ is set to high, confining the charge below SG, and the reference (pedestal) level is measured. In step 10, SG is driven to a potential higher than OG, transferring the charge packet to the floating SN, where the signal level is measured. At step 11, SG returns to a low potential. In step 12, OG is set to a potential lower than SN, pulling the charge packet back beneath SG.
Repeating steps 9 through 12 in a cycle allows the pixel’s charge packet to be measured an arbitrary number of times without degradation. This technique is known as Skipper measurement, and the total number of measurements taken is denoted by $N_S$.
Each of these measurements is calculated using the Dual Slope Integrator (DSI) technique. 
Multiple non-destructive measurements are averaged to reduce the readout noise to sub-electron levels. 
If charge $i$, measured at time $j$, has a mean value $Q_{i,j}$ with a standard deviation $\sigma_0$, the pixel value obtained by averaging $N_{S}$ Skipper measurements is given by

\begin{equation}
Q_i = \frac{1}{N_{S}} \sum_{j=0}^{N_{S}-1} Q_{i,j}
\label{eq:pix_average}
\end{equation}

This averaging process reduces the standard deviation of the readout noise \cite{cancelo2021low, Tiffenberg:2017aac} as in Eq. \ref{eq:sigma_reduction} due to the statistical independence of the measurements. Once the charge packet is ready to be removed, it is confined below the SN, finishing the readout sequence in step 11.

\section{Experimental Results}
\label{sec:experimental_results}
\begin{figure}[h]
    \centering
    \includegraphics[page=1,width=\linewidth, trim= {0 0cm 0cm 0}, angle=-90, clip]{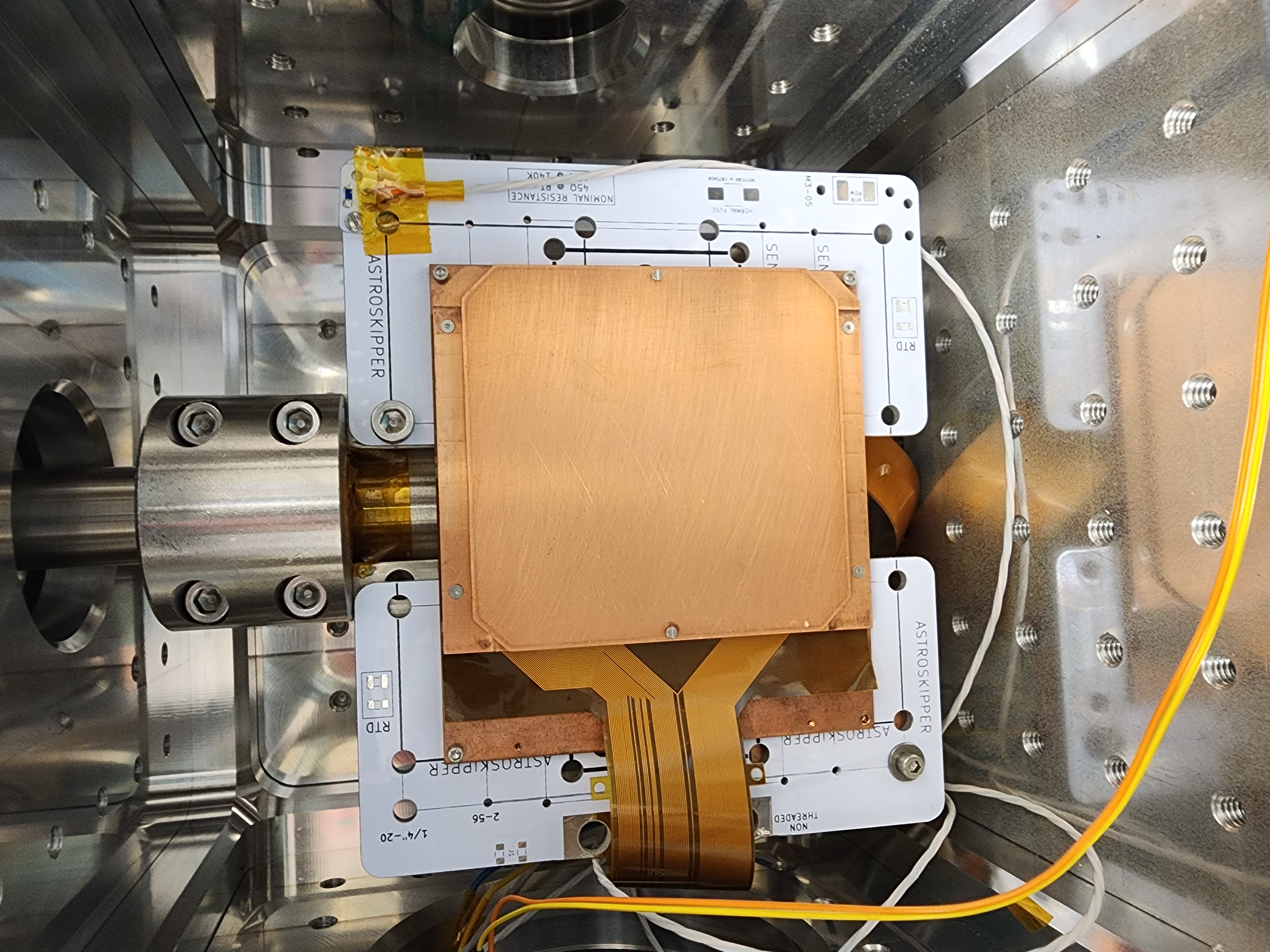}
    \caption{The photograph depicts the Skipper-CCD used to run the experiment, inside a vacuum chamber. The sensor is mounted on a cold plate that keeps it refrigerated at the operating temperature.}
    \label{fig:setup}
 \end{figure}

In this article, a CCD with skipper output stages was used for the measurements. The device features four amplifiers, one located on each corner of the sensor, with two pairs of two amplifiers connected through a common serial register, having one output stage at each end, as in the conceptual diagram explained in Section \ref{sec:charge_transfer_mechanisms}. The sensor is operated using the Low Threshold Acquisition controller with a desktop computer running the acquisition software to perform the tests \cite{cancelo2021low}. The CCD is housed inside a vacuum chamber, as shown in Figure \ref{fig:setup}. It is mounted on a copper plate thermally coupled to a cold finger, allowing stable operation at 140K. The cooling is provided with a Sunpower cryocooler mounted on top of the cube, while a Lakeshore controller is used to maintain a stable temperature. 

The disposition of this CCD and its use as a current source are described in the following subsection.
 ~%

\subsection{Current cell realization}
\begin{figure}[t]
    \centering
    \includegraphics[page=1,width=\linewidth, trim= {0 0 0 0}, clip]{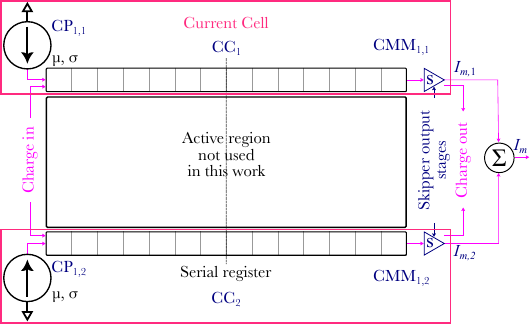}
    \caption{Schematic showing how to use a CCD as a current cell. The current source injects current into the serial register (left side), which is readout with a skipper output stage on the right side. Thereby, the current flows from left to right.}
    \label{fig:charge-injection-sketch}
 \end{figure}

A Skipper-CCD with four output channels was repurposed for two current cell configuration, where each half of the CCD acts as an independent cell, as illustrated in Figure \ref{fig:charge-injection-sketch}. Each half of the sensor has an output stage at each end, with the serial register located between them. The stages at one end of the serial register are configured for charge injection, which would form the CP of Figure \ref{fig:generic_architectures}, while those at the opposite end operate in standard readout mode, being the CMM from the block diagrams. The active region of the sensor is not used. 
Charge packets are sequentially injected and shifted through the serial register following the fill-and-spill procedure described in Section \ref{sec:charge_transfer_mechanisms} (Figure \ref{fig:charge-inj-sequence}), until each pixel of the serial register is filled with a non-empty charge packet. 
Charge injection and readout occur simultaneously at opposite ends of the serial register. This is made possible by using two different $V_{dr}$ voltages: one configured for the injection stage and another for the readout stage. An external power supply was used to set the $V_{dr}$ voltage of CMM$_{1,1}$ and CMM$_{1,2}$ to –21 V, while a dedicated clock controller operated the ohmic contact for CP$_{1,1}$ and CP$_{1,2}$.

Figure \ref{fig:charge-injection-sketch} illustrates the configuration diagram, showing two current sources at the left end of the CCD serial register, which correspond to the charge injection input stages. At the opposite end (right), the skipper output stages operate in their standard readout mode. In this configuration, the current can be measured as the rate of charge flow over time. At the end of the CC, the output currents $I_{m,1}$ and $I_{m,2}$ are added to obtain $I_{m}$.

 ~%

The results presented in this section correspond to two basic compact cells ($N_A=1$) in Figure \ref{fig:generic_architectures_a}, arranged in parallel ($N_C=2$), as shown in Figure \ref{fig:charge-injection-sketch}. 
At this stage, the charge (or equivalently, the current) generated by the different cells is physically summed at a common drain node, since all cells are connected to the same output potential. The quantities combined in post-processing correspond to the measurements provided by the CMMs. Among the architectures proposed in Section \ref{sec:multiple-cell_architecture_and_performance}, the selected implementation was chosen based on availability, and it enables operation with a single standard four-channel Skipper readout.

\subsection{Charge injection}
The charge-injection technique in Sec.\ref{sec:charge-injection-theory} is used to generate different charge-packet sizes by adjusting the potential $V_{dr}$ applied to the ohmic contact on the injection side. 

\begin{figure}[t]
    \centering
    \includegraphics[page=1,width=\linewidth, trim= {0 0 0 0}, clip]{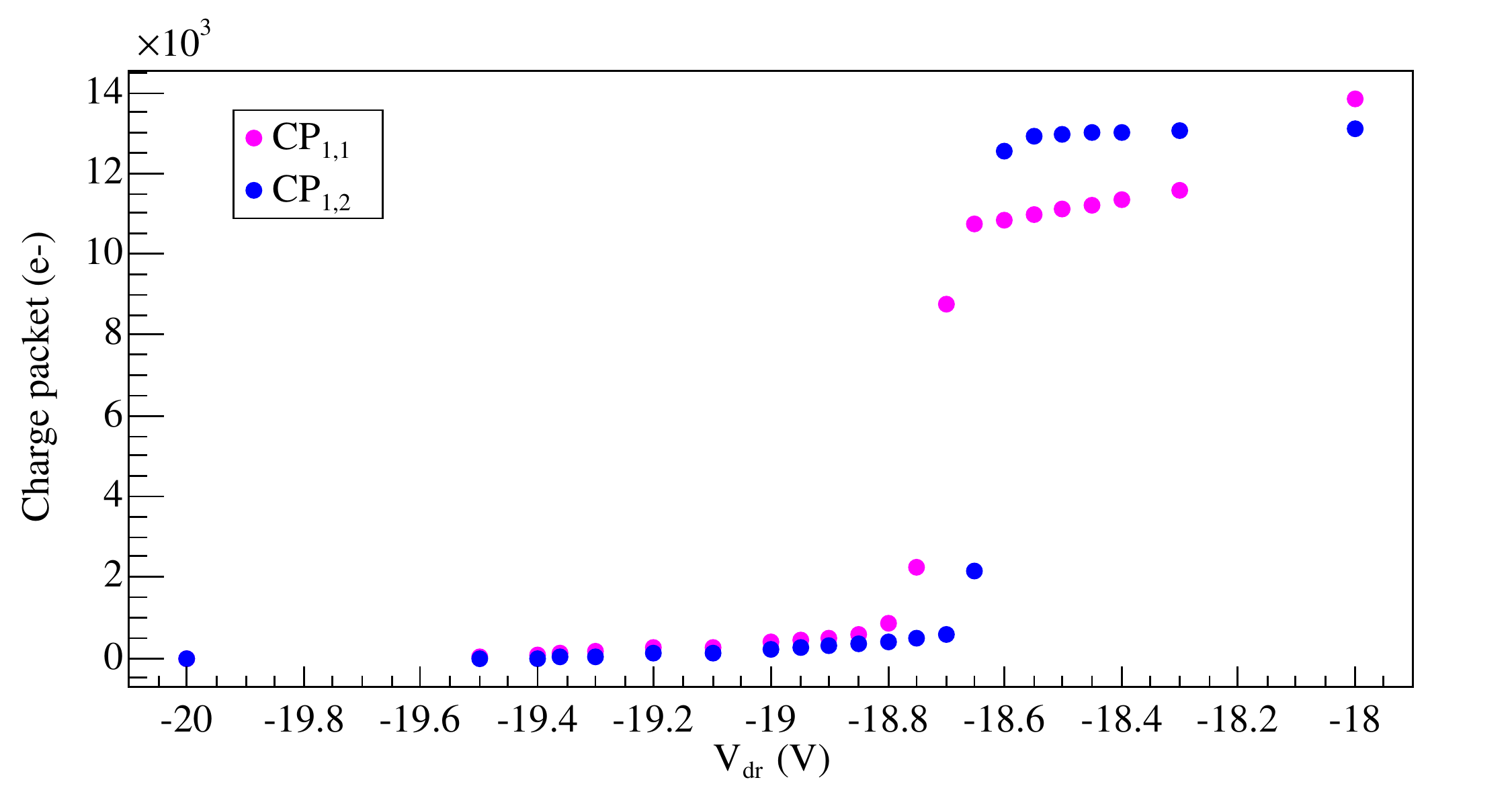}
    \caption{Measured injected charge as a function of the voltage applied to the ohmic contact $V_{dr}$. The charge injection response has three characteristic regions. No charge injected, low to moderate charge packets injected when the potential is similar to the channel, and surpassing this region, the charge floods to the pixels, reaching a plateau, given by the full well pixel capacity.}
    \label{fig:vdrain_vs_e}
 \end{figure}

Sources CP$_{1,1}$ and CP$_{1,2}$ from figure \ref{fig:charge-injection-sketch} were used to inject charge into their respective current cells, CC$_1$ and CC$_2$. Figure~\ref{fig:vdrain_vs_e} shows the injected charge, in units of electrons as a function of $V_{dr}$, with CP$_{1,1}$ in blue and CP$_{1,2}$ in magenta.
As $V_{dr}$ becomes less negative, the collected charge in the pixels increases and eventually reaches a plateau. The saturation levels of the two cells are similar, although a difference is observed between them.

As seen in Figure~\ref{fig:vdrain_vs_e}, for $V_{dr}$ $<-19.5$ V, no charge is injected because the potential remains below the channel potential.
In the second region, approximately between $-19.5 \text{ V} < V_{dr} < -18.8 \text{ V}$, the potential at the ohmic contact becomes comparable to the channel potential, producing a threshold-like effect that allows low to moderate charge packets to flow into the sense node as this threshold is surpassed. 
Finally, for higher $V_{dr}$ values, the potential exceeds the channel potential sufficiently to produce a charge-flood regime, where the collected charge reaches a plateau. The charge packets reached are around $10000\sim12000$ electrons depending on the CP and selected voltage.
This plateau corresponds to the maximum pixel capacity, the pixel full-well, which determines the largest charge packet a pixel can hold. No full-well optimization was performed in this work, and larger full-wells are possible in CCDs, as the $130$ke$^-$ previously described in Section \ref{sec:simulation_results}.
~%

\subsection{Charge readout and calibration}

\begin{figure}[t]
    \centering
    \includegraphics[page=1,width=\linewidth, trim= {0 0 0 0}, clip]{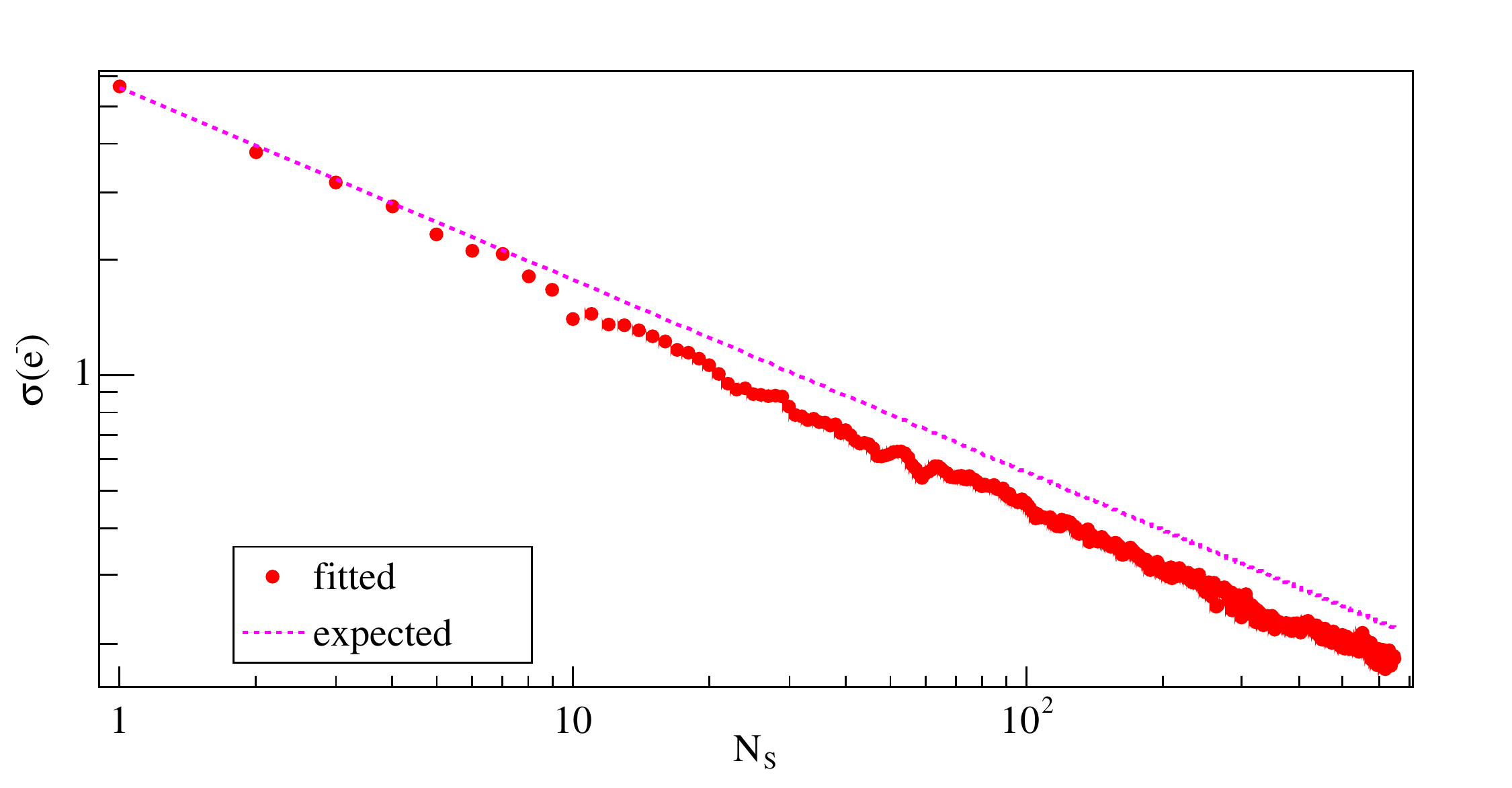}
    \caption{Readout standard deviation as a function of independent Skipper samples, $N_S$. The measured noise (red dots) decreases with $N_S$ and follows the expected $1/\sqrt{N_S}$ shown by the magenta dashed line.}
    \label{fig:noise-vs-nsamp}
 \end{figure}

The sensor performance is characterized in terms of readout noise as it is normally done for Skipper-CCD systems. For this, data was acquired using $N_S=650$ samples per pixel, for a $V_{dr}$ bias of $–19.37$ V (Figure~\ref{fig:vdrain_vs_e}). One of these datasets was used to estimate the readout noise from measurements of empty charge packets (i.e., with no charge injection). The readout standard deviation was then computed using all the $N_S$ values from 1 to 650. Figure~\ref{fig:noise-vs-nsamp} shows the resulting noise as a function of $N_S$. The red markers represent the measured standard deviation in electrons, while the magenta dashed line shows the expected $1/\sqrt{N_S}$ according to Eq. \ref{eq:sigma_reduction}. When plotted on logarithmic axes, the data closely follow this trend, confirming the expected noise reduction with increasing number of samples. The single-sample noise is slightly elevated due to correlated noise in the system. Increasing the number of samples decorrelates the noise, thereby reducing readout noise. For more details about noise decorrelation, refer to \cite{lapi2025sixteen}. The sensor performance when computing $N_S=650$ independent samples of the same charge packet is $\sigma=0.18$e$^-$.

 ~%
 \begin{figure}[h]
    \centering
    \includegraphics[page=1,width=\linewidth, trim= {0 0 0 0}, clip]{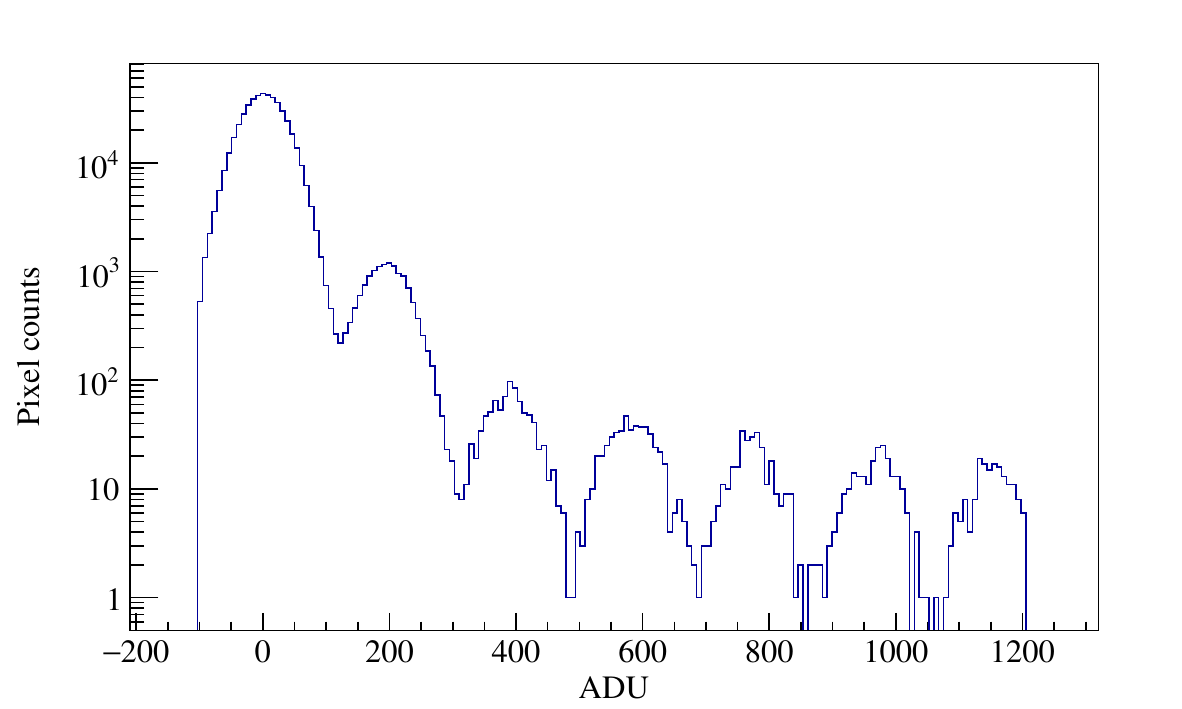}
    \caption{A fragment of the histogram of pixels obtained from the charge-injection data, showing counts in ADUs values. Multiple electron peaks are observed. The first distribution is the 0e- peak, and this fragment goes up to 6e-.}
    \label{fig:initial-peaks}
 \end{figure}~%
The same low-charge injection point was selected from Figure~\ref{fig:vdrain_vs_e} ($V_{dr}$= –19.37 V), and horizontal binning was employed to accumulate larger charge packets. In CCDs, binning consists of combining the charge from multiple adjacent pixels into a single readout pixel, allowing larger charge packets to be generated. In this case, as a low charge injecting point was selected, the large charge packets were achieved by adding the serial register pixels using binning.
Data was acquired using binning from 1 to 90, achieving electron-level resolution with $N_S = 650$. Through this configuration, charge packets up to 2500 e$^{-}$ were successfully generated and measured. Standard signal-processing steps for image sensors, such as crosstalk correction and overscan subtraction, to eliminate the offset of the distribution, were applied to the data. Figure~\ref{fig:initial-peaks} presents a fragment for the initial peaks of the histogram. The first peak corresponds to a 0e- value, the next to a 1e- value, and so on. By fitting a sum of gaussians to this distribution, the mean and standard deviation are obtained, and the mean value is then used as the ADU position of the electron peaks.
Once all the peaks have been processed, the electron values can be displayed as a function of the ADUs value.

\begin{figure*}[t]
    \centering

    \subfloat[]{
        \includegraphics[
            page=1,
            width=0.48\linewidth,
            trim={0 0 0 0},
            clip
        ]{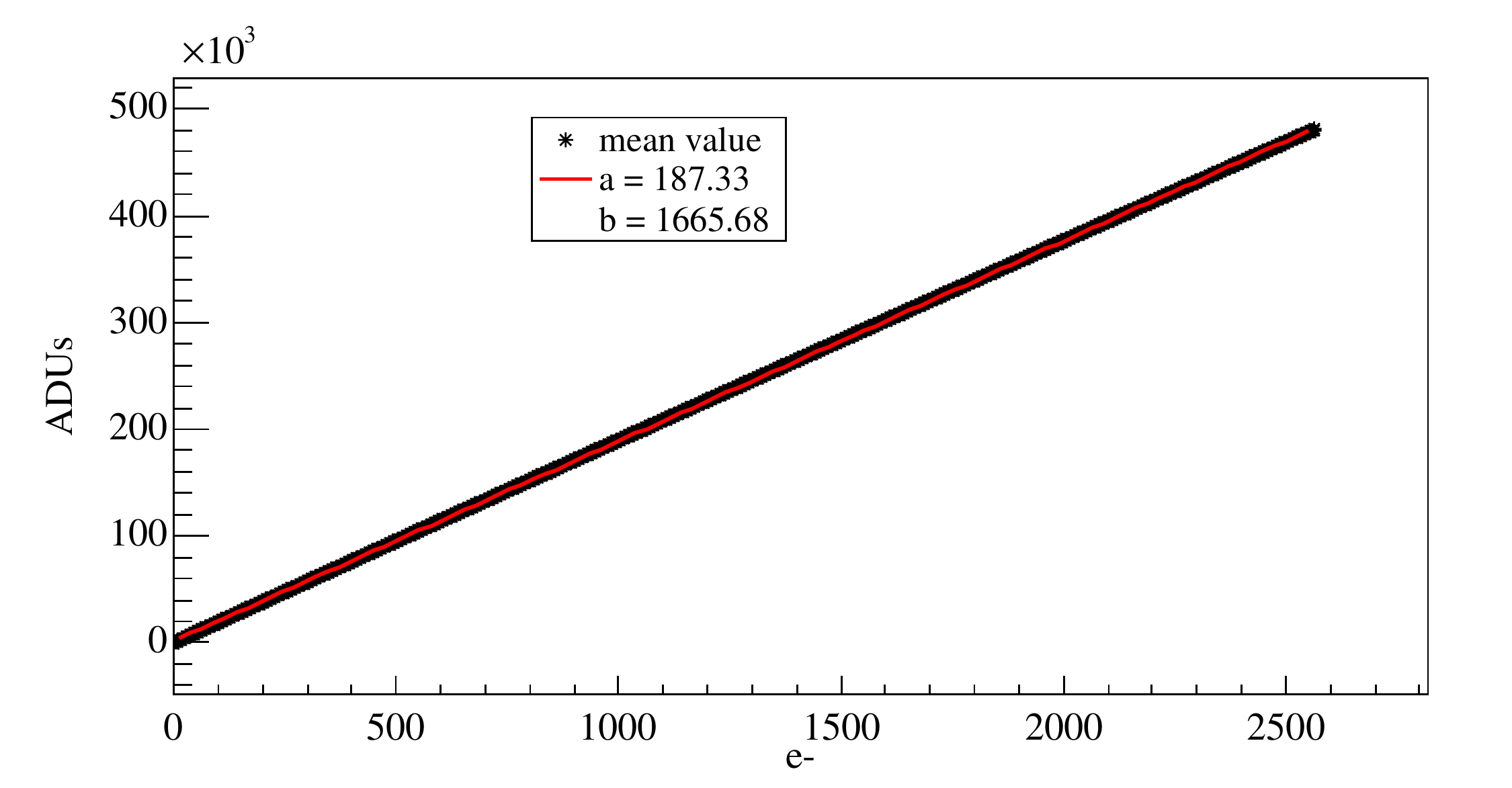}
        \label{fig:adu_vs_e_fit}
    }
    \hfill
    \subfloat[]{
        \includegraphics[
            page=5,
            width=0.48\linewidth,
            trim={0 0 0 0},
            clip
        ]{inl-dnl.pdf}
        \label{fig:fit_residuals}
    }

    \caption{Calibration linearity and the effect of a missed electron peak.
    (a) Mean ADU value of each electron-peak distribution as a function of
    the number of electrons. The linear regression is shown in red.
    (b) Difference between the data and the linear model, expressed in
    electron units. The solid line corresponds to the case in which all
    peaks are identified, whereas the dashed line corresponds to a
    calibration in which one peak was intentionally removed.}
    
    \label{fig:adu-vs-e-fit}
\end{figure*}

Figure~\ref{fig:adu-vs-e-fit} shows the mean ADU value of the electron peak distribution on the x-axis and the corresponding electron value on the y-axis. A linear regression is applied to this data, resulting in the relation $y= ax +b$, where $x$ is the ADU value and $y$ the corresponding electron value, with $a=0.0053, b=-8.878$.
The results may hide details related to linearity or potential missing electron counts during the fitting of the histogram peaks.
To evaluate potential missing peaks in the lookup table, the difference between the experimental data and the linear regression is computed.
The residual error is displayed in Figure~\ref{fig:fit_residuals}, the solid line corresponds to the case where all peaks were correctly identified.
In this case, the error curve is continuous and smooth, with a maximum deviation from the linear model in the full range of approximately 8e-. 
In contrast, deliberately removing one of the peak means from the lookup table produces a clear discontinuity in the residual curve at the location of the missing peak (blue line). Because the peak is not identified, the mean of the following distribution is assigned to the missing electron value, shifting the subsequent calibration by one electron and producing an error in the absolute electron count.

The overall variation, likely arising from the electronic amplification chain, does not affect the absolute calibration, since the method relies on counting the discrete electron multiplicities. A lookup table can therefore be implemented to directly map each electron number to its corresponding ADU value. This variation of around 2100 ADUs, which is equivalent to $\sim11.22$e-, is less than $0.5\%$ of the entire calibrated range, and can be neglected.

\begin{figure}[h]
    \centering
    \includegraphics[page=1,width=\linewidth, trim= {0 0 0 0}, clip]{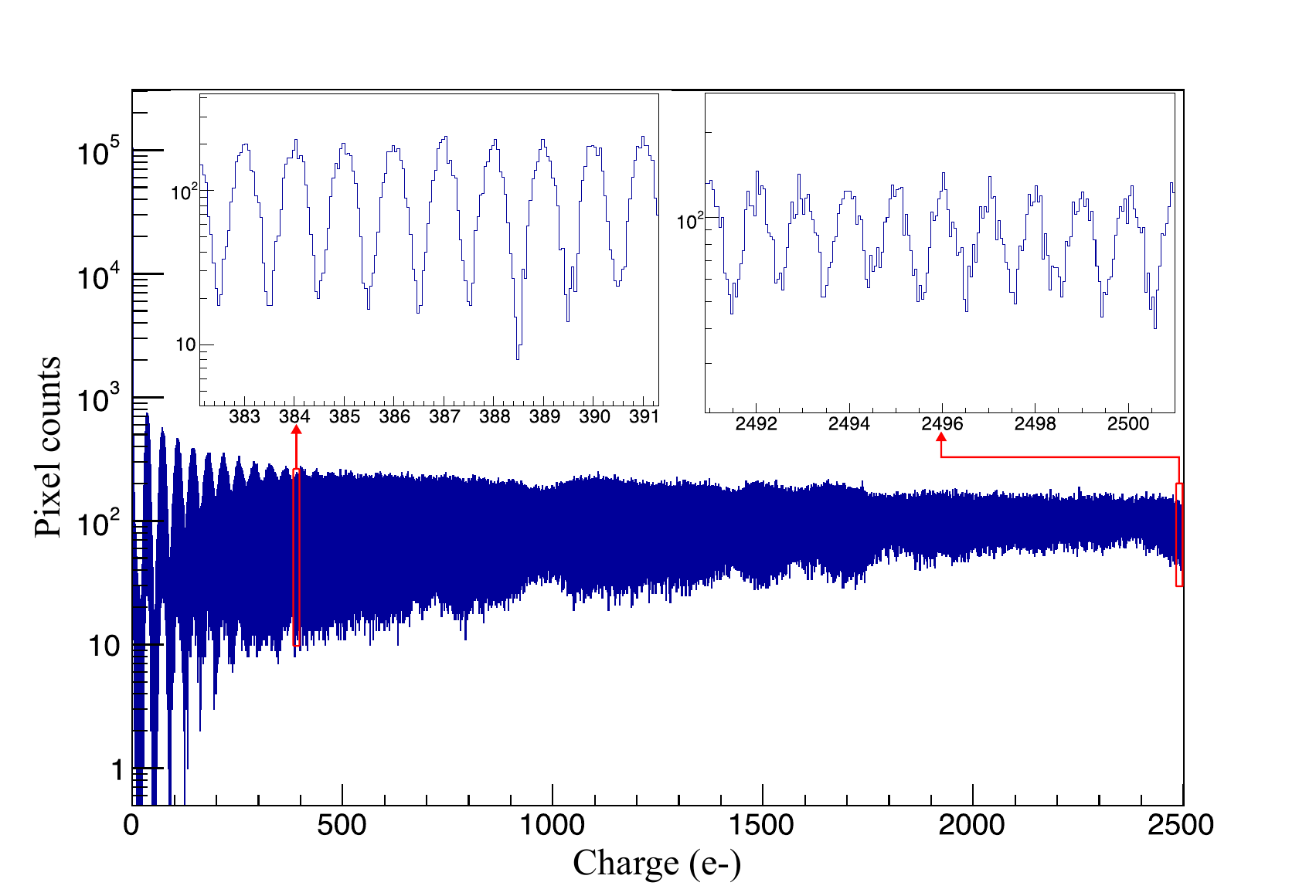}
    \caption{Complete histogram calibrated in electrons for a wide range. Two insets show the peak resolution for charge packets around 385e- and 2500e-.}
    \label{fig:histogram_full_calibration}
 \end{figure}

A representative histogram, including two insets, is shown in Figure~\ref{fig:histogram_full_calibration}.
The x-axis represents the electron count across a broad range (0–2500 e-), over which a complete ADU-to-electron calibration was performed. The discrete nature of the electron is observed across the entire range. Additionally, a normal distribution is fitted to the 0e- peak to measure the noise, yielding a standard deviation of 0.184e-. The bumps in the figure are due to the use of binning to collect the statistics.
~%

Achieving absolute charge calibration at the level of several thousand electrons is a capability not commonly available in conventional image sensors or other detectors, which typically rely on optical illumination and are limited by photon statistics and illumination stability. The method presented in this work enables a continuous and finely tunable range of charge packets to be generated without the use of light, relying instead on controlled charge injection combined with horizontal binning. This approach produces charge distributions with well-separated, uniformly spaced single-electron peaks, allowing precise calibration over a broad dynamic range using purely electronic means. To our knowledge, such an electrically generated, homogeneous single-electron charge spectrum, extending into the few-thousand-electron regime, has not been demonstrated previously, underscoring the significance and novelty of the technique.

\subsection{Current stability}
A long run was performed to explore the temporal stability of the current cell when operating on a minutes timescale. For this, a $V_{dr}$ potential of –18.5 V was selected from Fig.~\ref{fig:vdrain_vs_e}. At this operating value, a large amount of charge is injected, around $11100$e$^-$ and $13000$e$^-$ for both, CP$_{1,1}$ and CP$_{1,2}$, respectively. 

\begin{figure}[h]
    \centering
    \includegraphics[page=1,width=\linewidth, trim= {0 0 0 0}, clip]{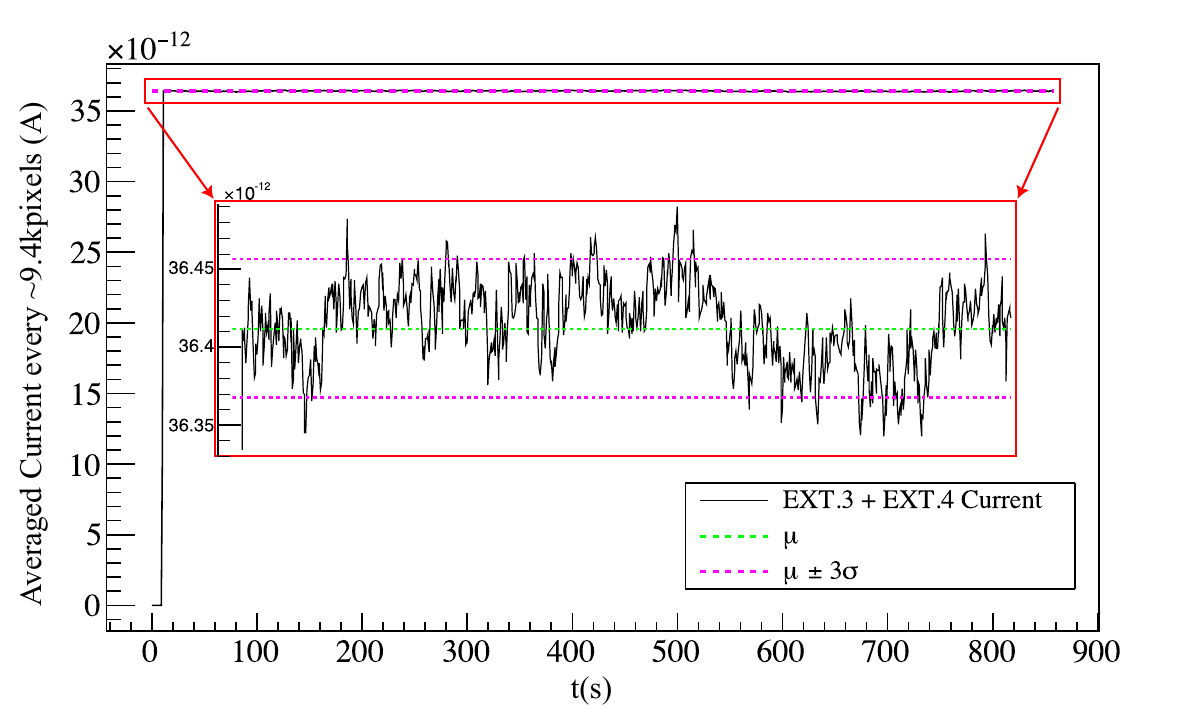}
    \caption{Total mean current per second, where the majority of the points are contained inside the 3$\sigma$ bounds. The inset reveals the current behavior during the $\approx15$ minutes readout. The current is obtained by measuring the charge packets, and normalizing by the readout time $T_{PIX}$.}
    \label{fig:current_over_time}
 \end{figure}

Fig.~\ref{fig:current_over_time} shows the total current over time, obtained by adding the measured charge with CMM$_{1,1}$ and CMM$_{1,2}$, and normalizing by $T_{PIX}$ rate, to compute the current in amperes. Each data point in the curve corresponds to the average current over a group of 9428 pixels, which approximately corresponds to one second of acquisition time. The inset highlights the green dashed line representing the mean current, while the magenta dashed lines indicate the $\mu \pm 3\sigma$ range.

As the current distribution is not known, 
to analyze the current stability and detect outliers, Chebyshev's inequality (Eq. \ref{eq:Chebyshev_inequality}) was used:

\begin{equation}
P(|X - \mu| \geq k\sigma) \leq \frac{1}{k^2}
\label{eq:Chebyshev_inequality}
\end{equation}

For this, the mean $\mu$ and standard deviation $\sigma$ were calculated using only the data from the first 60 seconds of acquisition. Then, the number of points that fell outside the bounds of $k\sigma$ (with $k=2,3$) was counted for the rest of the dataset (i.e., excluding the first 60 seconds). This approach allows us to test whether the behavior of the current remains statistically stable beyond the initial measurement window \cite{rabinovich2010evaluating}.

 ~%
 
The percentage of points that fell outside the $k\sigma$ limits was then compared to the upper bound predicted by Chebyshev’s inequality. For example, for $k = 3$, the inequality predicts that no more than $\frac{1}{9} \approx 11.1\%$ of the values should be outside the range $3\sigma$. The percentage observed in our data was significantly lower ($0.27\%$), confirming that the current signal is stable over time with minimal deviation.
\section{Basic cell layout}
\label{sec:layout_basic_cell}
\begin{figure}[h]
\centering
\includegraphics[page=1,width=\linewidth, trim={0 0 0 0}, clip]{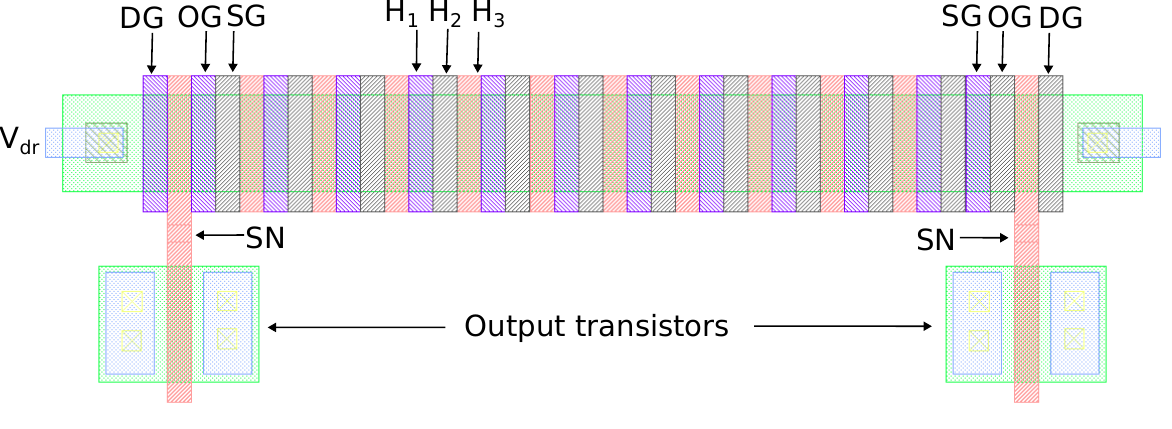}
\caption{Proposed design of an integrated circuit implementing the basic current-source cell.}
\label{fig:layout}
\end{figure}

Finally, we propose a complete implementation summarizing the studies previously shown in the diagram in Figure \ref{fig:generic_architectures_a}, in Section \ref{sec:charge-injection-theory}. 
The proposed cell has two skipper output amplifiers, one in each corner, connected by a ten-pixel serial register. This compact cell is easily scalable and cost-effective in terms of silicon area. Figure \ref{fig:layout} shows the layout of the proposed cell. It also indicated the main operating gates used in the charge transfer mechanisms, for both injection and readout, which were explained in Section \ref{sec:charge_transfer_mechanisms}. The basic current-source cell is designed as a compact and self-calibrating unit.

The cell architecture resembles a short serial register of a CCD, featuring ohmic contacts at both ends. Skipper amplifiers are added to both sides. As per the symmetry of the cell, any of the output amplifiers could be used as CP or CMM from Figure \ref{fig:generic_architectures_a}.
In the output side, the Skipper amplifier provides self-calibration capability and enables operation in the sub-electron noise regime. 
The amplifier on the charge-injection side serves to monitor the injected charge and the charge transport optimization, or as an extra measure unit. Having amplifiers at both ends of the analog memory aids in understanding how the charge transports. The amplifier on the injecting side could be used as an extra CMM to measure the injected packet, to later verify with the measurement of the output amplifier, in a feedback mode.
The readout signal is used in a feedback-control configuration, as illustrated in Figure~\ref{fig:objective_scheme}, where the reduced number of pixels in the serial register allows for a faster control response.

\section{Conclusions and future work}
\label{sec:conclusions}

To conclude, Sec.~\ref{sec:multiple-cell_architecture_and_performance} introduced scalable current cell architectures capable of transporting multiple electron packets. In Sec.~\ref{sec:simulation_results}, their signal, noise, and noise-to-signal ratio performance were analyzed for different configurations composed of one or more CMM blocks along the memory register. The effect of incorporating a quantization stage to reduce readout noise was evaluated. Although quantization is generally undesirable from a signal-processing perspective, it becomes advantageous here due to the discrete nature of the electron. Charge transport mechanisms for both injection and standard operation in the Skipper-CCD stage were also examined.

Experimentally, a Skipper-CCD with single-electron resolution was configured to implement one of the proposed architectures. The sensor response to charge injection was characterized, demonstrating charge packets up to $\sim$12,000e$^-$. Standard Skipper-CCD characterization measurements were performed. The combined use of the fill-and-spill method and binning enabled consistent generation of larger charge packets. Absolute calibration was achieved up to 2,500e$^-$, which, to the best of the authors’ knowledge, has not been previously reported. Finally, a compact layout design of the proposed current source architecture was presented.

As part of future work, we plan to experimentally validate the structure proposed in Section \ref{sec:layout_basic_cell} and characterize its performance. The next step will be to connect several of these cells in parallel, following the architecture shown in Figure \ref{fig:generic_architectures_c}. According to the simulations presented in Section \ref{sec:simulation_results}, this configuration is expected to increase the total output current.

\bibliographystyle{IEEEtran}
\bibliography{main.bib}

\end{document}